\documentclass[runningheads]{llncs}
\usepackage{graphicx}
\usepackage{amsmath}
\usepackage{amssymb}
\usepackage{bm}
\usepackage{mathtools}
\usepackage{xspace}
\usepackage[misc,geometry]{ifsym}
\begin{document}
\title{Implicit Neural Representations for Modeling of Abdominal Aortic Aneurysm Progression}
\titlerunning{Implicit Neural Representations for Aneurysm Progression}
%
\author{Dieuwertje Alblas\inst{1}\orcidID{0000-0002-7754-7405}\textsuperscript{(\Letter)} \and
Marieke Hofman \inst{1}\and
Christoph Brune \inst{1} \orcidID{0000-0003-0145-5069} \and
Kak Khee Yeung \inst{2, 3} \orcidID{0000-0002-8455-286X} \and
Jelmer M. Wolterink \inst{1} \orcidID{0000-0001-5505-475X}}

\authorrunning{D. Alblas et al.}
%
\institute{Department of Applied Mathematics, Technical Medical Centre, University of Twente, Enschede, The Netherlands \\ \email{d.alblas@utwente.nl} \and
Department of Surgery, Amsterdam UMC location Vrije Universiteit Amsterdam, Amsterdam, The Netherlands \and 
Amsterdam Cardiovascular Sciences, Microcirculation, Amsterdam, The Netherlands
}

\maketitle 

\begin{abstract}
Abdominal aortic aneurysms (AAAs) are progressive dilatations of the abdominal aorta that, if left untreated, can rupture with lethal consequences. Imaging-based patient monitoring is required to select patients eligible for surgical repair. In this work, we present a model based on implicit neural representations (INRs) to model AAA progression. We represent the AAA wall over time as the zero-level set of a signed distance function (SDF), estimated by a multilayer perception that operates on space and time. We optimize this INR using automatically extracted segmentation masks in longitudinal CT data. This network is conditioned on spatiotemporal coordinates and represents the AAA surface at any desired resolution at any moment in time. Using regularisation on spatial and temporal gradients of the SDF, we ensure proper interpolation of the AAA shape. We demonstrate the network's ability to produce AAA interpolations with average surface distances ranging between 0.72 and 2.52 mm from images acquired at highly irregular intervals. The results indicate that our model can accurately interpolate AAA shapes over time, with potential clinical value for a more personalised assessment of AAA progression.
\end{abstract}

\section{Introduction}
Abdominal aortic aneurysms (AAAs) are progressive local dilatations of the abdominal aorta of at least 30 mm that most frequently occur below the renal arteries. AAAs are mostly asymptomatic, but rupture of an AAA has a mortality rate of 70-80\% \cite{brewster2003guidelines}. To avert rupture, patients can undergo elective repair via either open surgery or an endovascular procedure. Patients become eligible for surgical repair if the diameter of the AAA exceeds a threshold (5.5 cm in men, 5.0 cm in women) or if the AAA diameter has increased more than 1 cm in a year~\cite{wanhainen2019editor}. 

Prior to elective repair, patients are monitored via periodic outpatient clinic visits and imaging with ultrasound or CT. Although these longitudinal images are primarily used to measure the diameter of the aneurysm, they contain a wealth of information that may be leveraged to better model AAA progression in individual patients~\cite{groeneveld2018systematic}. Detailed insight into personalised AAA progression has the potential to aid the physician in clinical decision-making by filling in the gaps in surveillance data. Previous efforts to model the progression of AAAs based on longitudinal imaging include models based on Gaussian processes that represent an underlying deformation field ~\cite{do2018prediction}, Markov chains~\cite{zhang2020intraluminal}, deep belief networks~\cite{jiang2020deep}, or CNNs operating on the surface of the AAA~\cite{kim2022deep}. 

Recently, implicit neural representations (INRs) have gained traction as natural representations for signals on a spatial or spatiotemporal domain~\cite{xie2022neural}. INRs are multilayer perceptrons that take continuous coordinates as input and output the value of the signal or function at that point~\cite{sitzmann2020implicit}. INRs are attractive representation models as derivatives of the signal can be analytically computed using automatic differentiation. In medical imaging, INRs have been used for, e.g., sparse-view CT reconstruction \cite{shen2022nerp,sun2021coil} and image registration~\cite{wolterink2022implicit}. Moreover, INRs can be used to accurately represent shapes~\cite{park2019deepsdf}, which has led to applications in cell shape synthesis~\cite{wiesner2022implicit} statistical shape modeling~\cite{ludke2022landmark,amiranashvili2022learning} or surface fitting based on point cloud annotations~\cite{alblas2022going}. 

In this work, we propose to use INRs with a time coordinate to represent a longitudinal 3D AAA model of a patient and investigate to what extent such a model can be used to \textit{interpolate} and \textit{extrapolate} the AAA surface in time.

 

\section{Methods}
We represent the evolving AAA surface as the zero level set of its temporal signed distance function (SDF). We parametrize this function by a neural network $f(\bm{x}, t; \theta)$, with weights $\theta$.

\subsection{Signed distance function}
A surface can be implicitly represented by the zero level set of its signed distance function. We consider a manifold evolving over time, that we represent by a temporal SDF: $SDF(\bm{x}, t): \mathbb{R}^3 \times \mathbb{R} \mapsto \mathbb{R}$. The value of the $SDF(\bm{x}, t)$ represents the minimum distance to the surface at location $\bm{x}$ at time $t$. The temporal SDF of an evolving 2D manifold $\mathcal{M}$ embedded in $\mathbb{R}^3 \times \mathbb{R}$ is defined as:
\begin{align}\label{def_SDF}
SDF_{\mathcal{M}}(\bm{x}, t) = \begin{cases}
- d(\bm{x}, \mathcal{M}) \xspace &\bm{x} \text{ inside } \mathcal{M} \text{ at time } t \\
0 & \bm{x} \text{ on } \mathcal{M} \text{ at time } t\\
d(\bm{x}, \mathcal{M}) & \bm{x} \text{ outside } \mathcal{M} \text{ at time } t.
\end{cases}
\end{align}
Moreover, the signed distance function is a solution to the Eikonal equation at each instance in time: $||\nabla_x SDF(\bm{x}, t)|| = 1, \forall \bm{x}, t$.

\subsection{Implicit Neural Representations}
\begin{figure}
    \centering
    \includegraphics[width=0.7\textwidth]{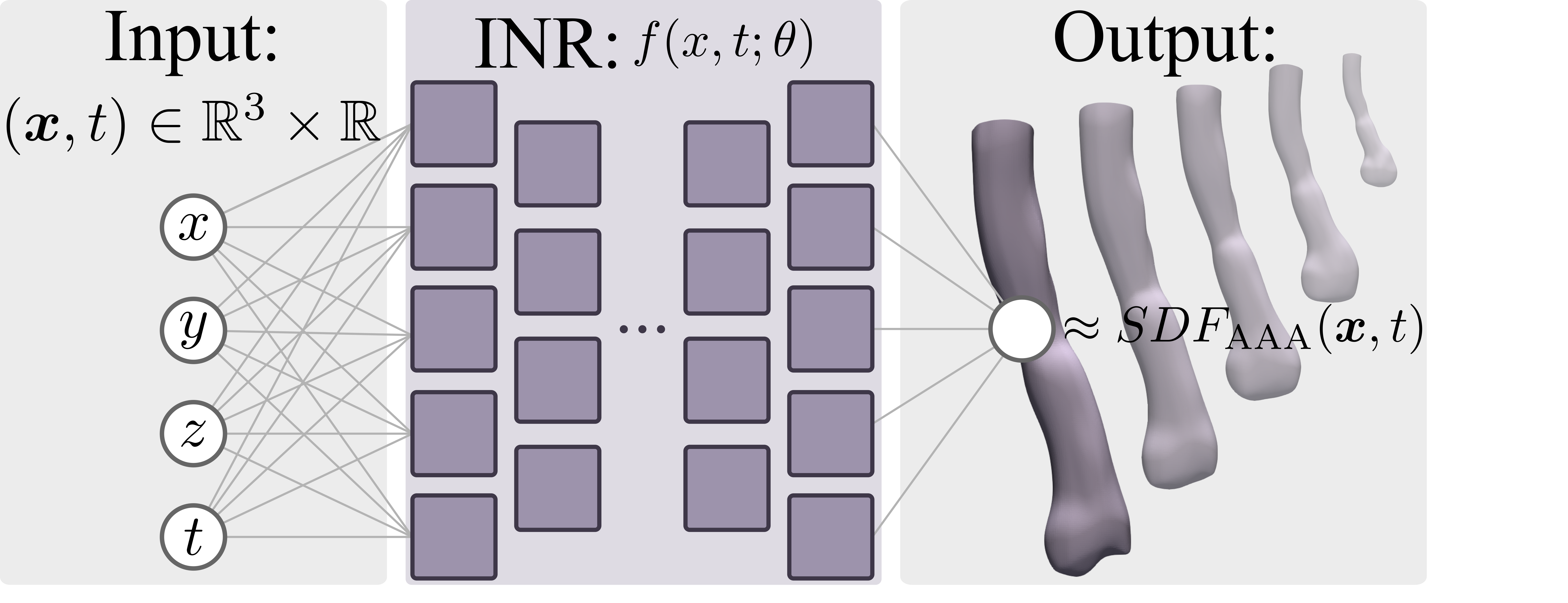}
    \caption{Schematic representation of our INR, taking spatiotemporal coordinates $(\bm{x}, t)$ as an input, outputting $SDF(\bm{x}, t)$ of the AAA surface. Note that a single INR represents the complete evolving AAA of a patient.}
    \label{fig:INR_scheme}
\end{figure}

In previous work, it has been shown that an SDF of a manifold can be represented by a neural network  ~\cite{alblas2022going,gropp2020implicit,park2019deepsdf,wiesner2022implicit}. Similarly, we embed the remodeling of the AAA over time in an implicit neural representation (INR). We use 4D coordinates from the spatiotemporal domain $\Omega: = [-1,1]^3 \times [-1, 1]$ as input to the network $f(\bm{x}, t; \theta)$. The output node of our INR approximates the SDF value at the input coordinate. Figure \ref{fig:INR_scheme} shows a schematic overview of our INR. 

We aim to reconstruct $SDF_{\text{AAA}}(\bm{x}, t)$ given a sequence of point clouds of the AAA surface, representing the aneurysm shape of a single patient over $J$ scans: $\{\mathcal{X}_{j}\}_{j=1,...,J}$, where $\mathcal{X}_{j,\cdot} \subset [-1, 1]^3$. We denote individual points on the $j^{\text{th}}$ AAA surface $\bm{x}_i^j$. To optimise the INR, we sample points on and off the AAA surface at multiple instances in time.\\

The loss function we use to optimize the INR consists of two terms: a term $\mathcal{L}_{\text{data}_j}$ at each time point $t_j$ where we have ground-truth scan data, and a term $\mathcal{L}_{\text{reg}}$ that regularises the SDF at times the surface is unknown. 
\begin{align}
\mathcal{L}(\theta) &= \sum\limits_{1 \leq j \leq J}  \mathcal{L}_{\text{data}_j}(\theta) + \mathcal{L}_{\text{reg}}(\theta),\\
\begin{split}
\mathcal{L}_{\text{data}_j}(\theta) &= \frac{1}{N_j}\sum\limits_{1 \leq i \leq N_j} |f(\bm{x}_i^j, t_j;\theta)| + \lambda_1 \mathbb{E}(||\nabla_x f(\bm{x}, t_j; \theta)|| - 1) ^2 \\ & + \lambda_2 \mathbb{E}(|\nabla_t f(\bm{x}, t_j; \theta)|) \end{split}\\
\mathcal{L}_{\text{reg}} & = \lambda_3 \mathbb{E} \left(||\nabla_x f(\bm{x}, \tilde{t}; \theta)|| - 1 \right)^2 +  \lambda_4 \mathbb{E}\left(|\nabla_t f(\bm{x}, t; \theta)| \right).
\label{eq:reg}
\end{align}
The first term of $\mathcal{L}_{\text{data}_j}$ was introduced in \cite{gropp2020implicit}. It ensures $SDF(\bm{x}_i^j, t_j) = 0$ for all points $\bm{x}_i^j$ in pointcloud $\mathcal{X}_j$, i.e. that points that are known to be on the AAA surface are indeed on the zero level set of the SDF. The remaining terms in both parts of the loss function regularise the INR's spatial and temporal gradient. As these terms do not depend on pointcloud data, we evaluate them both at times $t_j$ as well as times data is unavailable. Regularising the norm of the spatial gradient was also introduced in \cite{gropp2020implicit} and enforces the INR to be a solution to the Eikonal equation. We evaluate this term at time $t_j$ in $\mathcal{L}_{\text{data}_j}$, and at an arbitrary time point $\tilde{t}$ in $\mathcal{L}_\text{reg}$. The temporal regularisation term is introduced in this work to restrict temporal changes of the INR. These are evaluated at time $t_j$ and at multiple arbitrary time points in $\mathcal{L}_{\text{data}_j}$ and $\mathcal{L}_{\text{reg}}$ respectively. 

\begin{figure}[t!]
    \centering
    \includegraphics[width=\textwidth]{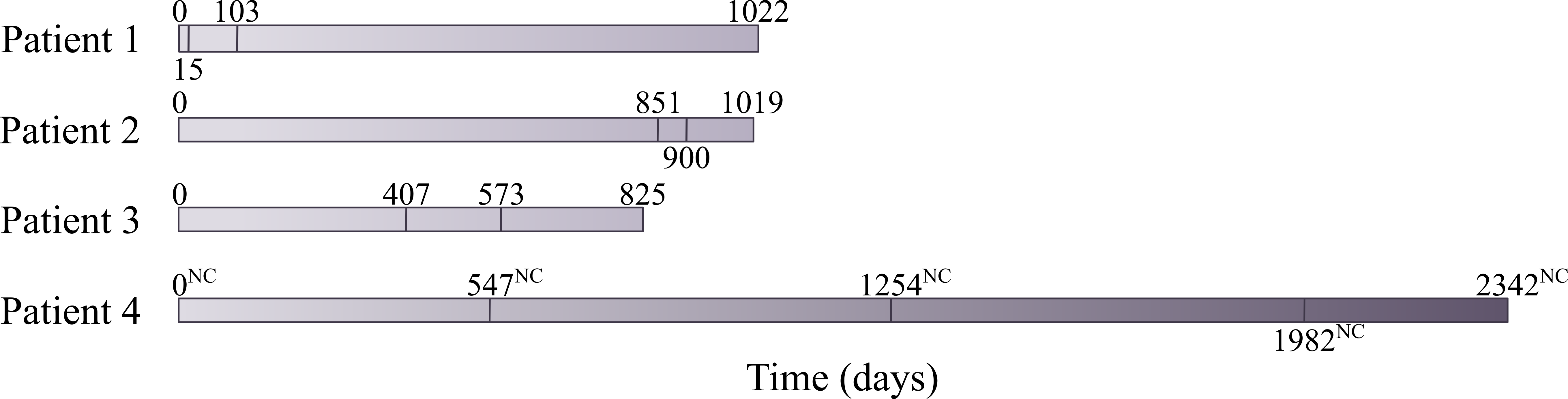}
    \caption{Timeline of the CT scans of the four patients with longitudinal data, showing scan instances in days. Non-contrast scans are indicated with \textsuperscript{NC}.}
    \label{fig:timeline}
\end{figure}

\subsection{Data}
We retrospectively included longitudinal CT data of four patients scanned at Amsterdam AMC (Amsterdam, The Netherlands) between 2011 and 2020. Three patients were scanned four times, and one patient was scanned five times (Fig.~\ref{fig:timeline}). Scan dates were shifted so that the first scan date of each patient became day 0. Patient 1 was scanned three times between day 0 and day 103, followed by a gap of almost three years. The first follow-up image of Patient 2 was after 851 days, after which two additional follow-up images were acquired relatively soon. Patient 3 was scanned more regularly. The follow-up for Patient 4 is the longest, with over 75 months of follow-up. CT scans were a mixture of non-contrast and contrast-enhanced images.

We obtained automatic segmentations of the AAA and vertebra in each of these patients. All CT scans were processed using TotalSegmentator~\cite{wasserthal2022totalsegmentator}, a Python library based on nn-UNet~\cite{isensee2021nnu} that segments $>100$ structures in 3D CT images. This library segmented the vertebra with good accuracy in both non-contrast and contrast-enhanced images and the AAA with high accuracy in all non-contrast images. However, segmentation of the AAA in contrast-enhanced images was unsatisfactory. Instead, we used an in-house dataset of 80 contrast-enhanced CT images of AAA patients with annotations of the AAA ranging between the top of the T12 vertebra and the iliac bifurcation to train an additional nn-UNet model. This model achieved a mean Dice similarity coefficient of 0.90 on a separate test set consisting of 13 contrast-enhanced CT scans.

\subsection{Preprocessing}
In order to evaluate local changes in shape over time, all shapes should be aligned in the same coordinate system. For this, we used rigid registration in ITK on the vertebra segmentations~\cite{do2018prediction}. Subsequently, the surface of each aorta was extracted from the mask and represented as a point set. This resulted in aligned pointcloud representations of the AAA surface for each scan. Finally, before serving as input to the network, the spatial coordinates of the pointclouds of each patient were jointly normalized to the $[-1,1]^3$ domain. Similarly, the time scale of each patient was normalized to the $[-1,1]$ interval.

\begin{figure}[t!]
\includegraphics[width=\textwidth]{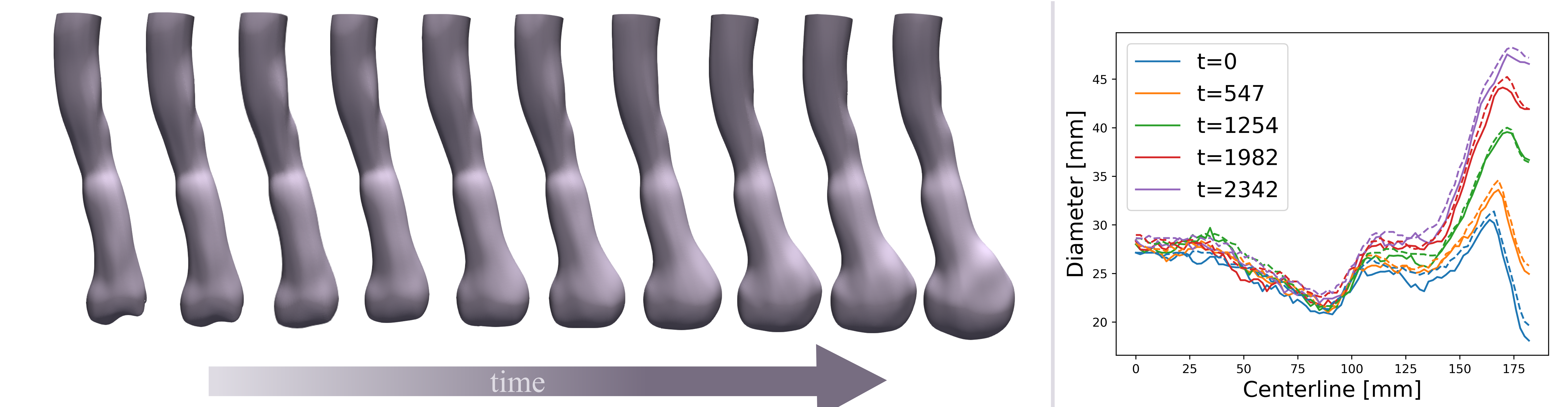}
\caption{An optimised INR can be used to extract shape interpolations at an arbitrary number of time points, here we show results for Patient 4. \textit{Left:} We show extracted shapes at ten regularly spaced intervals in time. \textit{Right:} Diameter plots along the centerlines of the aorta, comparing the ground-truth segmentation mask (solid) to the surface fitted by the network at five time points where reference CT scans are available (dashed).}
\label{fig:tensteps}
\end{figure}

\begin{figure}[ht!]
    \centering
    \includegraphics[width=\textwidth]{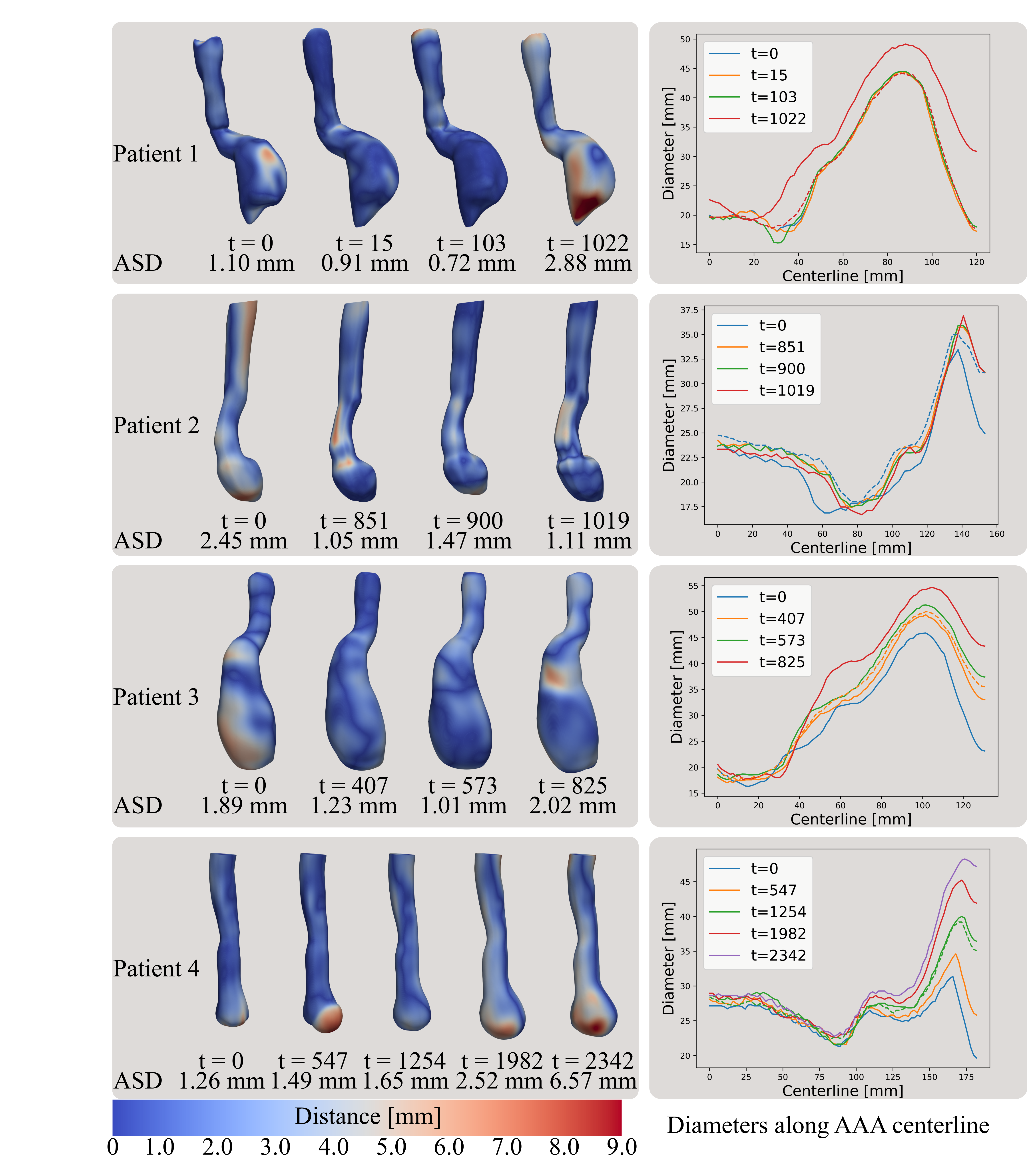}
    \caption{\textit{Left:} Inter- and extrapolated AAA shapes for each scan. Colors indicate distances to reference shapes, averages are indicated below each AAA. \textit{Right:} Diameter profiles along each aorta. Solid lines represent reference diameters, dashed lines show interpolated or extrapolated diameters. A complete set of diameter plots can be found in the Appendix.}
    \label{fig:overview_interpolations}
\end{figure}

\section{Experiments and Results}
In all cases, we used an MLP with six fully connected layers containing 256 nodes with ReLU activations and a final node representing the estimated SDF of the AAA surface. Like \cite{alblas2022going,gropp2020implicit}, we used a skip connection, connecting the input to the third hidden layer. The regularization coefficients were set to $\lambda_1 = \lambda_2 = \lambda_3 = \lambda_4 = 0.1$. We used an Adam optimizer with a learning rate of 0.0001 to train our network for 25,000 epochs on an NVIDIA Quadro RTX 6000 GPU. The batch sizes depended on the size of point clouds and ranged between 2877 and 6027. 

\subsection{Interpolation and Extrapolation}
For each patient, we first optimised a single INR (Fig. \ref{fig:INR_scheme}) based on point clouds from all available scans. Because the spatiotemporal input coordinates to the INR are continuous, we can retrieve a shape at any point in time at any resolution. We visualize this in Fig. \ref{fig:tensteps}(\textit{left}), where we show ten AAA shapes of Patient 4 at regularly spaced intervals. In Fig. \ref{fig:tensteps}(\textit{right}) we compare the diameters along the AAA centerlines of the ground-truth segmentation masks to the AAA surfaces reconstructed by the network, represented by solid and dashed lines respectively. We observe that the model accurately represents the AAA shapes at scan instances and will thus be used to evaluate the next experiment.

Next, we performed a series of leave-one-out experiments in which we optimised an INR for a patient but left out one of the time points. We used the optimised INR to estimate what the surface would have been at that time point, and compare it to the real reconstruction. 
Figure \ref{fig:overview_interpolations} shows the results of this leave-one-out experiment. Results for individual patients are visualized per row. The left column in each row shows the interpolated or extrapolated AAA shapes predicted by the network when that scan was left out of the training data. Colors indicate the minimal surface distance between the interpolated AAA surface and the reference AAA surface, where lower is better. The right column contains diameter plots for each aorta along its centerline estimated based on an inscribed sphere method \cite{gharahi2015growth,jiang2020deep}. Solid lines represent the diameters of the reference AAA surfaces, and the dashed line represents the diameter of the AAA from a scan that was left out. Note that we here show the diameter profile for one leave-one-out experiment per patient and that a full set of diameter profiles can be found in the Appendix.

Figure \ref{fig:overview_interpolations} shows that the INR model can \textit{interpolate} AAA shapes to a decent extent. For example, in Patient 3, the interpolated surfaces at $t=407$ and $t=573$ had average surface distances of 1.23 and 1.01 mm, respectively, compared to the ground-truth shapes. This is also reflected in the diameter plot for Patient 3, where the interpolated (dashed) line for $t=407$ days closely follows the reference (solid) line. The results in Figure \ref{fig:overview_interpolations} also indicate that interpolation might work better in cases where the interval between scans is shorter. For example, interpolation results for Patient 1 at $t=15$, which is only 15 and 88 days apart from two other scans, have an average surface distance of 0.91 mm. In contrast, interpolation results for Patient 4 at $t=547$, which is 547 and 707 days apart from two other scans, show relatively large errors on the aneurysm sac. However, this is not consistently the case. For Patient 2, the ASD is 1.47 mm when interpolating at $t=900$ days, which is larger than the ASD when interpolating for $t=851$ days. From the diameter plots shown for Patients 3 and 4, we see that interpolations of the model consistently lie between the surrounding two scans and are close to the diameters of the reference shape. 


Results also indicate that extrapolation is challenging for the model. The INR particularly struggles to extrapolate over bigger time gaps. For Patient 1, we observe that the extrapolations at $t=0$ days and $t=1022$ days have worse results than the interpolations. Moreover, the extrapolation at $t=1022$ days differs more from the reference shape than at $t=0$ days due to the difference in time gaps. The diameter profiles for Patient 1 and Patient 2 reveal that the model tends to reconstruct the surface of the last known shape. We hypothesize that this might be due to the temporal regularization term in Eq. \ref{eq:reg}.

Finally, Figure \ref{fig:overview_interpolations} indicates that our INR model reacts strongly to small misalignments of the original AAA shapes. Following \cite{do2018prediction}, we register AAA shapes based on segmentation masks of the vertebrae, but this alignment might lead to small local shifts of the AAA. For example, the result for Patient 2, $t=0$ in Fig. \ref{fig:overview_interpolations} shows errors on the healthy part of the aorta, an area that, in principle, should not show growth over time. 

\section{Discussion and Conclusion}
In this work, we have obtained a personalised model for AAA progression, based on longitudinal CT data. We combine fully automatic state-of-the-art image segmentation methods, registration, and shape modeling with implicit neural representations and adequate regularisation terms to build personalised models of an evolving anatomical structure. In experiments with four longitudinally scanned AAA patients, we have demonstrated how the model represents the evolving shape of an AAA over time. This may impact patient monitoring and treatment; accurate knowledge about the progression of an AAA allows the physician to personalise surveillance and time intervention better based on AAA diameter and growth rate~\cite{wanhainen2019editor}.

One appealing aspect of our approach is the continuity of the implicit neural representations. This allows us to reconstruct an AAA mesh at any point in time, at any desired resolution. We have here modeled shape changes over multiple years with sparse and irregularly spaced shape data. Modeling this change through linear interpolation of alternative surface representations, such as meshes or point clouds, would require point-to-point correspondence, a challenging problem that we here circumvent. Moreover, since our network relies on pointcloud data, it is agnostic to imaging modality. This is important for longitudinal studies of AAAs, where imaging modalities such as MRI and 3D US are increasingly used. All these scans can be incorporated into this framework as long as we can extract AAA surfaces. Furthermore, because we represent an evolving shape in space and time in a differentiable neural network, we can add any gradient-based regularisation term to the loss function. We have here included an Eikonal term and temporal regularization, but this framework could be further extended. 
Lastly, we found that our model is sensitive to errors in the initial alignment of AAA shapes. Although we have followed~\cite{do2018prediction} in registering based on the location of the vertebrae, better results can likely be achieved by registering based on other landmarks, such as the renal arteries and iliac bifurcation.


One limitation of the current approach is the relatively limited test set of four longitudinally scanned patients, which we aim to increase in future work. Moreover, our approach is purely based on morphology and does not include other biomarkers for AAA growth and rupture~\cite{groeneveld2018systematic}. In future work, we will investigate if we can incorporate, e.g., results of computational fluid dynamics and fluid-structure interaction modeling in our model~\cite{kim2022deep}. Furthermore, additional optimization constraints could more properly model the pathophysiology of aneurysms. Whereas our temporal regularization term now aimed to minimize the gradient of the SDF, in future work, we could optimize this gradient within biologically plausible growth rates. This kind of regularization could also be obtained in a data-driven way, by learning a generalisable model from a larger set of patients with longitudinal data. Finally, there is evidence that intraluminal thrombus shape plays a key role in AAA remodeling \cite{zhang2020intraluminal}, and it might be beneficial to explicitly represent thrombus in our INR~\cite{alblas2022going}.



In conclusion, we have shown that INRs are promising tools in modeling AAA evolution. In future work, this flexible model could be extended with biologically plausible regularization terms and hemodynamic parameters.

\section{Acknowledgements}
Jelmer M. Wolterink was supported by the NWO domain Applied and Engineering Sciences VENI grant (18192).

%
%
\bibliographystyle{splncs04}
\bibliography{literature}

\newpage 

\appendix
\section{}
\label{sec:appendix}

\begin{figure}[h]
\centering
\includegraphics[width=\textwidth]{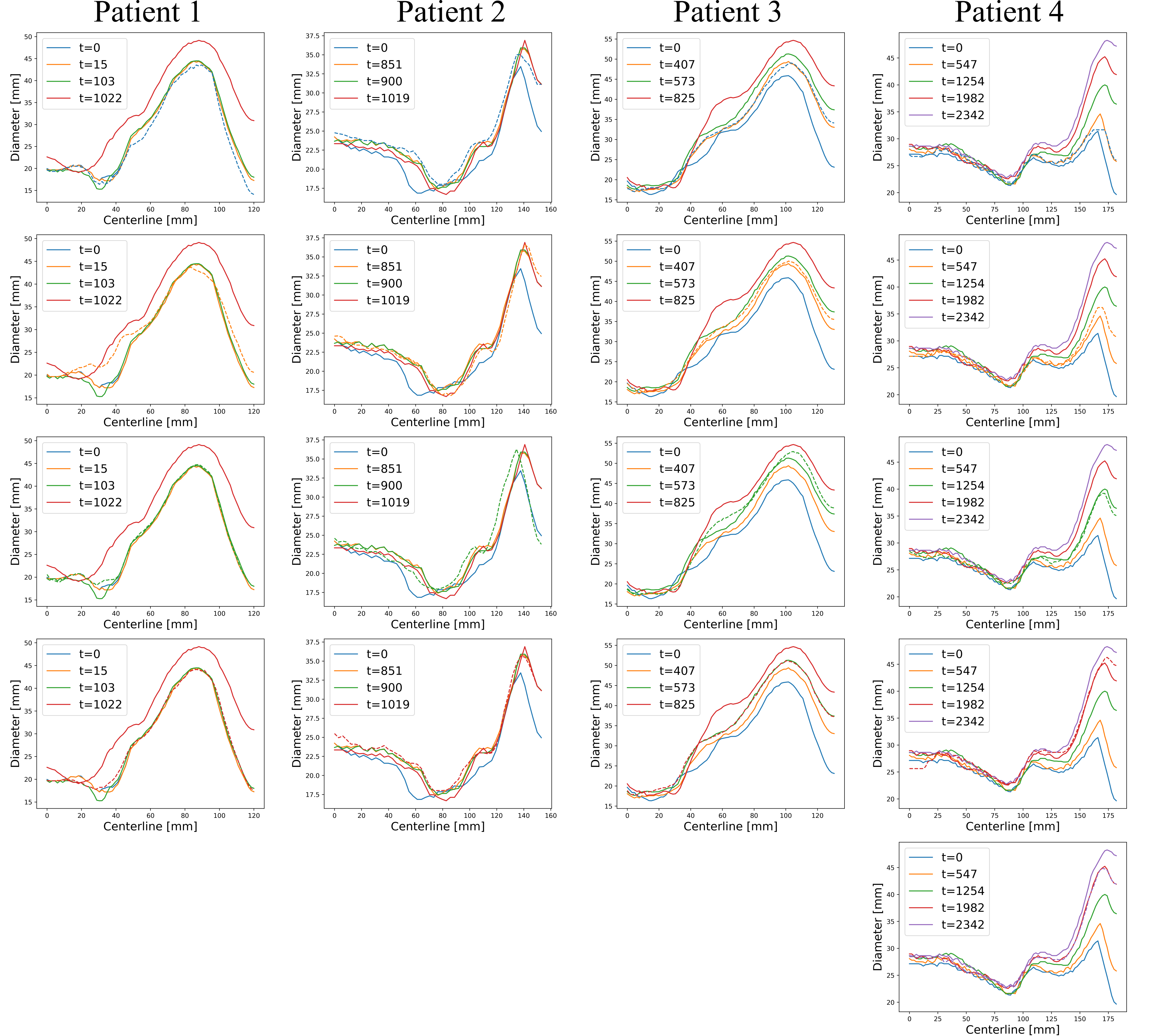}
\caption{Complete overview of the diameter plots resulting from the series of leave-one-out experiments. The solid lines represent the diameters of the reference AAA along the centerline, the dashed lines represent the diameters of the scan that was left out of the training data.}
\label{fig:overview_diameters}
\end{figure}

\end{document}